\documentclass[manuscript,nonacm]{acmart} 
\usepackage{soul}
\usepackage{hyperref}
\usepackage{hyperxmp}
\usepackage{graphicx,comment}
\usepackage{amsmath}
\usepackage{algorithm}
\usepackage{verbatim}
\usepackage{multirow}
\usepackage{booktabs}
\usepackage{makecell}
\usepackage{longtable}
\usepackage{soul}
\usepackage{enumitem}
\usepackage{xspace}
\usepackage{bm}
\usepackage{longtable}
\usepackage{caption}
\usepackage{adjustbox}

\usepackage{color, colortbl}
\definecolor{codegreen}{rgb}{0,0.6,0}
\definecolor{codegray}{rgb}{0.5,0.5,0.5}
\definecolor{codepurple}{rgb}{0.58,0,0.82}
\definecolor{backcolour}{rgb}{0.95,0.95,0.92}
\definecolor{LightCyan}{rgb}{0.88,1,1}
\definecolor{LightRed}{RGB}{255, 204, 203}

\usepackage{amsmath}


\makeatletter
\def\adl@drawiv#1#2#3{%
        \hskip.5\tabcolsep
        \xleaders#3{#2.5\@tempdimb #1{1}#2.5\@tempdimb}%
                #2\z@ plus1fil minus1fil\relax
        \hskip.5\tabcolsep}
\newcommand{\cdashlinelr}[1]{%
  \noalign{\vskip\aboverulesep
           \global\let\@dashdrawstore\adl@draw
           \global\let\adl@draw\adl@drawiv}
  \cdashline{#1}
  \noalign{\global\let\adl@draw\@dashdrawstore
           \vskip\belowrulesep}}
\makeatother

\usepackage{pifont}

\usepackage{minibox}
\newcounter{observcntr}

\usepackage{array}
\newcolumntype{L}[1]{>{\raggedright\arraybackslash}p{#1}}

\AtBeginDocument{%
  \providecommand\BibTeX{{%
    \normalfont B\kern-0.5em{\scshape i\kern-0.25em b}\kern-0.8em\TeX}}}

\copyrightyear{2023}
\acmYear{2023}
\setcopyright{rightsretained}
\acmConference[WDC '23]{The 2nd Workshop on the security implications of Deepfakes and Cheapfakes}{July 10, 2023}{Melbourne, VIC, Australia}
\acmBooktitle{The 2nd Workshop on the security implications of Deepfakes and Cheapfakes (WDC '23), July 10, 2023, Melbourne, VIC, Australia}
\acmDOI{10.1145/3595353.3595882}
\acmISBN{979-8-4007-0203-7/23/07}

\begin{document}

\title{From Promise to Peril: Rethinking Cybersecurity Red and Blue Teaming in the Age of LLMs}


\author{Alsharif Abuadbba}
\affiliation{%
  \institution{CSIRO's Data61}
  \country{Australia}}
\email{sharif.abuadbba@data61.csiro.au}

\author{Chris Hicks}
\affiliation{%
  \institution{The Alan Turing Institute}
  \country{United Kingdom}}
\email{c.hicks@turing.ac.uk}

\author{Kristen Moore}
\affiliation{%
  \institution{CSIRO's Data61}
  \country{Australia}}
\email{kristen.moore@data61.csiro.au}

\author{Vasilios Mavroudis}
\affiliation{%
  \institution{The Alan Turing Institute}
  \country{United Kingdom}}
\email{vmavroudis@turing.ac.uk}

\author{Burak Hasircioglu}
\affiliation{%
  \institution{The Alan Turing Institute}
  \country{United Kingdom}}
\email{bhasircioglu@turing.ac.uk}

\author{Diksha Goel}
\affiliation{%
  \institution{CSIRO's Data61}
  \country{Australia}}
\email{diksha.goel@data61.csiro.au}

\author{Piers jennings}
\affiliation{%
  \institution{Loughborough University}
  \country{United Kingdom}}
\email{p.jennings@lboro.ac.uk}






\renewcommand{\shortauthors}{Alsharif Abuadbba et al.}



\begin{abstract}


Large Language Models (LLMs) are poised to transform the cybersecurity landscape by augmenting both offensive (red team) and defensive (blue team) operations. With the capacity to automate and enhance tasks such as threat detection, intelligence synthesis, adversary simulation, and incident response, LLMs promise a new era of efficiency and scalability. Red teams can leverage LLMs to plan attacks, craft phishing content, simulate adversarial behaviors, and generate exploit code, while blue teams can deploy them to aggregate threat intelligence, assist with root cause analysis, and streamline security documentation. However, these capabilities come with significant caveats.

This position paper examines the implications of LLMs across key cybersecurity frameworks, including the MITRE ATT\&CK and NIST Cybersecurity Framework (CSF), offering a structured analysis of where and how LLMs intersect with critical red and blue team functions. We explore both the strengths and limitations of current LLM capabilities, emphasizing the urgent need for governance, standardization, and real-world evaluation benchmarks.

Our analysis indicates that while LLMs demonstrate fluency, versatility, and utility across cybersecurity tasks, they remain brittle in high-stakes, context-rich environments. Limitations such as constrained context retention, hallucinations, reasoning deficiencies, and prompt sensitivity undermine their reliability in complex operational settings. Furthermore, integrating LLMs into real-world workflows might introduce significant concerns—including dual-use risks, adversarial misuse, and the erosion of human oversight. Malicious actors can potentially exploit these models to rapidly scale cyber operations, automate reconnaissance, and obfuscate attack paths, thereby lowering the technical barrier to launching sophisticated threats.
To guide responsible and safer adoption, we outline strategic recommendations that include maintaining human-in-the-loop oversight, improving explainability, employing privacy-aware  and secure integration practices, and building systems resilient to adversarial use. As organizations move toward AI-enabled cyber defence, a nuanced understanding of LLMs’ risks and operational impacts is essential to ensure these tools enhance, rather than compromise, cybersecurity posture.



\end{abstract}

\maketitle
\section{Introduction}

Large Language Models (LLMs) are rapidly becoming pivotal tools in cybersecurity. 
Their ability to interpret and generate code, reason over complex inputs, and operate across natural language interfaces has opened new frontiers for both attackers and defenders. Once restricted to resource-rich organizations, LLMs are now widely accessible, making advanced cyber capabilities available at unprecedented scale. 
This position paper examines the implications of this shift for red and blue teaming operations, exploring both the opportunities LLMs create and the risks they introduce.

Among the areas where LLMs are already having visible impact is in \textit{red Vs blue teaming}--a practice central to cybersecurity preparedness. In these adversarial exercises, red teams emulate real-world attackers, probing for vulnerabilities by simulating tactics, techniques, and procedures (TTPs)~\cite{redteaming, kovavcevic2020red, kouremetis2025occult}.  The goal is not simply to break in, but to rigorously test defenses and provide actionable feedback to improve detection and response capabilities. Blue teams, meanwhile, are responsible for defending infrastructure: monitoring systems, assessing risks, analyzing threats, and implementing hardening and recovery measures. These roles often align with widely adopted frameworks such as the NIST Cybersecurity Framework (CSF)~\cite{nistCyberFramework} and MITRE ATT\&CK~\cite{mitre}.


As interest in LLMs accelerates, both red and blue teams have begun integrating them into their workflows. Red teams are experimenting with LLMs to automate tasks such as phishing campaigns, generating exploit code, simulating adversaries, and supporting reconnaissance. On the defensive side, blue teams are exploring LLMs for threat intelligence synthesis, incident documentation, and root cause analysis. These early use cases demonstrate the models' promise, but also their limitations. Most current applications remain narrow, focused on language-based tasks, and fall short in addressing the broader, data-intensive, and action-oriented demands of many cybersecurity operations—such as lateral movement, dynamic response, and real-time adaptation.

In this position paper, we explore the evolving role of LLMs in cybersecurity through the lens of red and blue teaming. We examine their contributions and constraints across practical scenarios, map their alignment with established frameworks such as MITRE ATT\&CK and the NIST CSF, and identify challenges and risks that remain underexplored. Our aim is to guide a responsible and strategic integration of LLMs, one that enhances cybersecurity posture without eroding human oversight, operational resilience, or ethical safeguards.

\section{Motivation and Threat Landscape}
The cybersecurity threat landscape in 2024 reached unprecedented intensity, marked by the accelerating volume, velocity, and sophistication of attacks targeting global infrastructure. A synthesis of legal analysis and data projections from 2024–2025 estimates cybercrime costs at \$9.5 trillion in 2024, with forecasts exceeding \$10.5 trillion by 2025~\cite{mohsin2025cybercrime}. High-profile incidents such as the Ticketmaster breach—impacting 560 million users—illustrate the scale and societal impact of these threats. Additionally, Amazon’s Chief Security Officer, CJ Moses, reported (Wall Street Journal, Nov 2024) that Amazon now faces nearly 1 billion cyber threats per day—a dramatic surge attributed in part to the growing use of artificial intelligence in both offensive operations and defensive needs~\cite{wlj2025amazon}.

This rapidly evolving threat environment demands a fundamental rethinking of cybersecurity paradigms. In particular, the convergence of LLMs with red and blue teaming introduces powerful new capabilities—alongside novel risks. As LLMs are increasingly integrated into both offensive tools and defensive frameworks, understanding their dual-use nature becomes imperative. This requires re-examining conventional red teaming methodologies, anticipating emerging attack surfaces, and proactively addressing the challenges, risks, and governance considerations associated with their adoption. A nuanced understanding of these dynamics is critical to building secure and resilient systems in the LLM age.

\section{Foundations: LLMs and Cybersecurity Operations}\label{sec:background}
To understand how LLMs may be safely and effectively integrated into cybersecurity workflows, we first provide a brief overview of how LLMs function and how red and blue team operations are structured.

\subsection{LLMs from First Principles}

LLMs are built on the transformer architecture~\cite{vaswani2017attention}, a foundational breakthrough in deep learning that enables models to process and generate natural language with remarkable fluency. These models are pretrained on vast corpora of text using autoregressive objectives~\cite{radford2019language}, predicting the next word in a sequence, which allows them to capture complex statistical patterns in language, logic, and even code. 


To make them useful in practical settings, LLMs are often fine-tuned on domain-specific tasks~\cite{devlin2018bert} and aligned with human intent through techniques such as reinforcement learning from human feedback (RLHF)~\cite{ziegler2019fine}. 
These post-training strategies improve performance and usability but do not eliminate core limitations. They play a critical role in improving safety, particularly in high-stakes applications like cybersecurity or bioterrorism prevention. They also reflect broader research efforts to manage trade-offs in model capability, safety, and scalability, as highlighted by emerging studies on architectural limits and scaling laws~\cite{kaplan2020scaling}.

\begin{figure}
    \centering
    \includegraphics[width=0.9\textwidth]{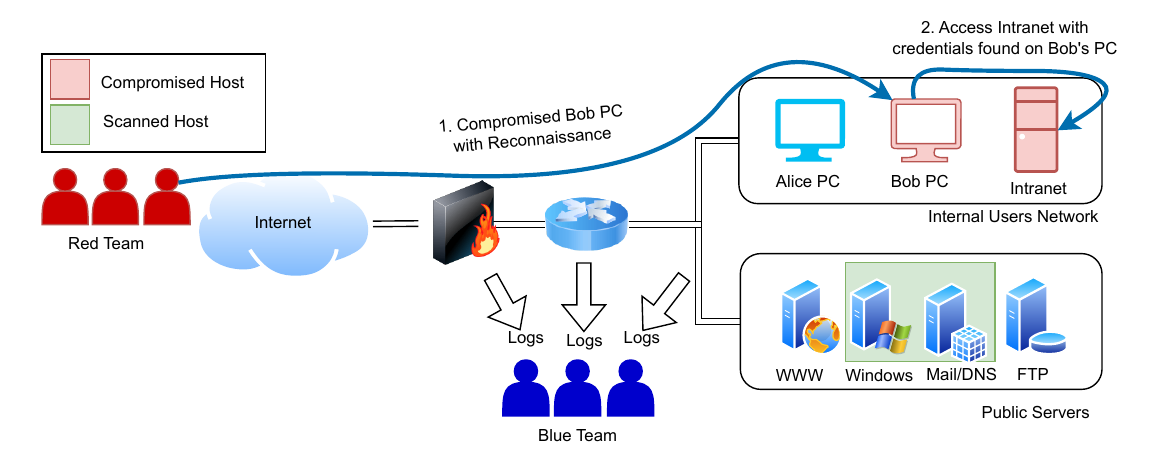} 
    \caption{An illustrative scenario of red team intrusion and corresponding blue team response pathways.}
    \label{fig:redblueteaming}
\end{figure}

\subsection{Red and Blue Team Key Activities}
Figure~\ref{fig:redblueteaming} presents an overview and illustrative example of the roles and activities involved in red and blue teaming. We next outline the common operational activities of red and blue teams. 
\subsubsection{Common Red Team Activities}
We utilise the MITRE ATT\&CK framework as a comprehensive reference for adversary tactics, techniques, and procedures derived from real-world observations. This framework has been widely adopted as a foundation for red teaming and threat modeling across various sectors, including government, academia, and industry~\cite{mitre,al2024mitre}. As illustrated in Table~\ref{tab:mitre}, the framework categorizes adversarial behavior into  distinct tactics, organized as rows. In this position paper, we highlight seven representative tactics (out of 14 on MITRE ATT\&CK framework) to demonstrate the varying levels of difficulty in automating these tasks using LLMs in Section~\ref{sec:strength-ch}.

\begin{table}[h!]
    \centering
    \captionsetup{justification=centering}
    \caption{Simplified MITRE ATT\&CK Red Teaming Framework with Examples.}
    \label{tab:mitre}
    \begin{adjustbox}{max width=\textwidth}
    \begin{tabular}{l|l|l}
        \hline
        \textbf{Tactic} & \textbf{Techniques} & \textbf{Illustration/Example} \\
        \hline
        Reconnaissance & Active Scanning & Scanning open ports using tools like Nmap \\
                       & Phishing for Information & Sending spear-phishing emails with malicious attachments \\
                       & Search Open Technical Databases & Crawling Shodan for exposed devices \\
        \hline
        Privilege Escalation & Abuse Elevation Control Mechanism & Exploiting Windows UAC to gain higher privileges \\
                            & Access Token Manipulation & Token impersonation using tools like Mimikatz \\
                            & Create or Modify System Process & Modifying a system service to execute malware \\
        \hline
        Defense Evasion & Abuse Elevation Control Mechanism & Disabling antivirus using administrative privileges \\
                       & Debugger Evasion & Detecting and avoiding debugging environments \\
                       & Deobfuscate/Decode Files or Info & Decrypting encoded payloads to execute malicious code \\
        \hline
        Credential Access & Adversary-in-the-Middle & DNS spoofing to intercept credentials \\
                         & Brute Force & Repeatedly attempting passwords against a service \\
                         & Steal Application Access Token & Extracting OAuth tokens from a compromised device \\
        \hline
        Lateral Movement & Exploitation of Remote Services & Exploiting RDP or SMB to move laterally \\
                        & Lateral Tool Transfer & Moving malicious binaries to a remote host \\
                        & Remote Services & Gaining control of remote systems via SSH or RDP \\
        \hline
        Command and Control & Application Layer Protocol & Using HTTP/S for communication with a C2 server \\
                           & Content Injection & Injecting malicious scripts into web applications \\
                           & Traffic Signaling & Using DNS tunneling for covert communication \\
        \hline
        Exfiltration & Automated Exfiltration & Uploading data to a cloud storage service \\
                    & Exfiltration Over Alternative Protocol & Using FTP instead of HTTP to exfiltrate data \\
                    & Exfiltration Over Web Service & Using Google Drive or Dropbox to exfiltrate files \\
        \hline
    \end{tabular}
    \end{adjustbox}
\end{table}

\subsubsection{Common Blue Team Activities}

We adopt the NIST CyberSecurity Framework (CSF) 2.0~\cite{nistCyberFramework}, a widely adopted approach for cybersecurity risk management, which complements the MITRE ATT\&CK framework with a set of security controls used by blue teams to counter threats and mitigate risks. As shown in Table~\ref{tab:csf}, the CSF organises blue team activities into six core functions: govern, identify, protect, detect, respond, and recover. All of the CSF functions relate to one another, with govern informing the implementation of the other core functions and referred to in Section~\ref{sec:strength-ch}.

\begin{table}[h!]
    \renewcommand{\arraystretch}{1.6}
    \centering
    \captionsetup{justification=centering}
    \caption{NIST CSF Core Functions with Example Risk and Security Controls.}
    \label{tab:csf}
    \begin{adjustbox}{max width=\textwidth}
    \begin{tabular}{l|L{0.23\textwidth}|L{0.58\textwidth}}
        \hline
        \textbf{Core Function} & \textbf{Purpose} & \textbf{Example Risk Management Activities and Security Controls} \\
        \hline
        Govern &
        Establish and monitor risk management strategy, expectations, and policy &
        \parbox[t]{\linewidth}{
            Understand legal, regulatory, and contractual requirements \\
            Determine expected capabilities, outcomes, and services \\
            Standardise methods for calculating and prioritising risks
        } \\
        \hline
        Identify &
        Understand current cybersecurity risks &
        \parbox[t]{\linewidth}{
            Maintain an accurate model of authorised network communication \\
            Identify vulnerabilities and threats to assets, people, and processes \\
            Establish, maintain, and communicate incident response plans
        } \\
        \hline
        Protect &
        Use security controls to manage risks to assets &
        \parbox[t]{\linewidth}{
            Maintain a policy defining access permissions and authorisations \\
            Establish and apply configuration management policies \\
            Create, protect, maintain, and test data backups
        } \\
        \hline
        Detect &
        Find and analyse possible attacks and compromises &
        \parbox[t]{\linewidth}{
            Monitor networks and assets for potential adverse events \\
            Analyse potential incidents or suspicious behavior \\
            Integrate cyber threat intelligence
        } \\
        \hline
        Respond &
        React to detected cybersecurity incidents &
        \parbox[t]{\linewidth}{
            Execute incident response plans\\
            Perform root cause analysis\\
            Stakeholder incident notification and information sharing
        } \\
        \hline
        Recover &
        Restore assets and operations after an incident &
        \parbox[t]{\linewidth}{
            Select, scope, prioritise, and perform recovery actions \\
            Verify the integrity of restored assets, systems, and services \\
            Communicate progress in restoring capabilities
        } \\
        \hline
    \end{tabular}
    \end{adjustbox}
\end{table}

\section{Strengths, Challenges, and Ethical Implications}\label{sec:strength-ch}




\subsection{Emergent Strengths of LLMs in Cybersecurity Operations} 
LLMs are beginning to demonstrate tangible benefits across cybersecurity operations. Here we highlight key areas where LLMs are enhancing the effectiveness and efficiency of red and blue teams, drawing on recent empirical findings and practical use cases.

\subsubsection{Enhanced Red Teaming Efficiency}
LLMs have enhanced red team operations' in cybersecurity across various TTPs~\cite{deng2024pentestgpt}. 
(1) Planning Support: LLMs can analyze tasks, devise plans, and recommend actions such as executing penetration testing tools or generating executable code to achieve specific objectives~\cite{shashwat2024preliminary}.
(2) Execution Support: With appropriate guidance, LLMs assist in conducting introductory-level penetration testing, particularly in targeted tasks. For instance, they can perform website hacking techniques such as SQL injection and cross-site scripting~\cite{fang2024llm}, execute and evaluate LLM-generated commands for network threat testing~\cite{moskal2023llms}, and exploit known Common Vulnerabilities and Exposures (CVEs) when provided with instructions~\cite{fangel2024llmX}. 
(3) External Tool Integration: LLMs can interact with external tools, such as web search engines, enhancing the breadth and depth of information accessible during testing.
(4) Rapid Iteration: LLMs can iteratively refine their actions based on feedback from previous outputs, improving effectiveness over time with human assistance.
(5) Automated Reporting: LLMs expedite time-consuming tasks like summarizing findings and documentation, streamlining the reporting process and allowing Red Teams to focus on more complex aspects of penetration testing.


\subsubsection{Enhanced Blue Teaming Efficiency}   

Blue teams spend a large proportion of time collecting, parsing, and documenting information about security risks~\cite{lavigne_14_visSOC,brown15_threat_mgmt,ullah_secanaly19}, but compared to red team activities there is relatively little research documenting the effectiveness of LLMs in blue team operations~\cite{wang2024shieldgpt}. Nevertheless, LLMs have the potential to greatly enhance blue team operations across a range of activities~\cite{wu2024threatmodeling}. We illustrate this by identifying five key LLM capabilities with the potential to productively augment one or more of the NIST CSF~\cite{nistCyberFramework} core functions introduced in Section~\ref{sec:background}. From each of these capabilities, LLMs can: (1) Information Aggregation and Analysis: assist in parsing lengthy legal, regulatory, and contractual frameworks to extract mutual risk management requirements (govern), as well as aggregate cyber threat intelligence from diverse sources during the identification phase (identify). (2) Automated Documentation and Reporting: facilitate the drafting of information-sharing communications and incident notifications for stakeholders in response to security incidents (respond). (3) Policy and Configuration Management: support the development and ongoing maintenance of asset and network configuration policies (protect). (4) Decision support: act as co-pilots alongside blue team members, aiding in the execution of incident response plans (respond). (5) Process Automation: write scripts to automate tasks such as backup creation and validation (protect and recover).

\subsection{Key Technical Challenges of LLMs in Cybersecurity Operations}
Technical challenges caveating current LLM capabilities include context length limitations, prompt brittleness, hallucinations, behavioral alignment, and unreliable evaluation practices~\cite{kaddour2023challengesllm, laskar-etal-2024-systematic,wang2025large}. Here, we contextualize these challenges from the perspective of red and blue team cyber operations. 

\subsubsection{Short term memory (i.e., context length)} Despite advancements in context length, LLMs continue to struggle with efficiently processing and maintaining focus on relevant parts of the input~\cite{hsieh2024ruler}. This limitation poses several challenges in cybersecurity operations, where contextual awareness is crucial for effective decision making. 
The most significant issue arising from context limitations is the loss of context over time. During long interactions (i.e., long token sequences) LLMs may fail to recall important information from earlier stages, leading to misaligned outputs or degraded performance. For red teamers, this is especially problematic in multi-stage attack simulations involving tactics such as \textit{Reconnaissance (Phishing)} Table~\ref{tab:mitre}, where tracking the progression of an attack through various phases requires consistent memory of prior steps. Similarly, blue teams relying on LLM-based \textit{detection} tools, Table~\ref{tab:csf}, may miss threats that are distributed across large volumes of benign  data due to the model’s inability to maintain relevant information across long contexts.
One notable consequence of this limitation is task repetition. When LLMs cannot remember prior interactions, they may redundantly reanalyze previously completed tasks or suggest actions that have already been taken. This may have directly deleterious effects (e.g., encrypting data twice could have consequences ranging from wasting resources to permanent denial of access) and could also reduce effectiveness, particularly in red teaming scenarios like simulating \textit{Command and Control (Content Injection)}, where an LLM-based system might fail to recognize that certain injection techniques have already been attempted (i.e., increasing the risk of detection). Blue teams might encounter compounding inefficiencies as both the execution and documentation of incident response procedures become prone to redundant steps.



\subsubsection{Hallucinations.}  When a language model generates false or misleading information with high confidence, particularly when it lacks access to verified or current data, it is said to be suffering from a hallucination. This poses significant challenges for red and blue teaming, where the consequences of acting on false information may be particularly severe. 
First, inaccurate outputs and false positives can mislead security teams---for example, an LLM might fabricate an attack path involving \textit{Tactic (Persistence)  and Technique (Hijack Execution Flow)}  when no such exploit is feasible, leading to unnecessary mitigation efforts. Second, undetected errors and flawed conclusions can skew threat models and result in poor security recommendations. In the context of \textit{redential Access (Exploitation for Credential Access)}, Table~\ref{tab:mitre}, an LLM may suggest fictitious credentials as attack vectors, diverting attention from actual threats. Hallucinations in blue team incident reporting could hinder or subvert downstream recovery, incident response planning, and stakeholder notification activities. Without robust, real-time validation mechanisms, these hallucinations can undermine the effectiveness of cyber operations and decision making.



\subsubsection{Reasoning limitations} 
Despite their fluency, LLMs often lack robust ``reasoning'', especially in multi-step logic, contextual adaptation, and strategic decision-making, limiting their effectiveness across red and blue team activities. In particular: (1) Challenges in End-to-End Automation:
LLMs struggle to coordinate multi-stage attack sequences, affecting coherence across phases like reconnaissance, exploitation, lateral movement, and exfiltration. For instance, during reconnaissance (\textit{Active Scanning}), poor prioritization may lead to inefficient exploitation in Remote Services, disrupting the attack chain (Table~\ref{tab:mitre}). Similarly, their inability to adapt tactics undermines adversary emulation in Phishing, where nuanced, context-aware actions are critical.
On the defensive side, limited reasoning during root cause analysis may lead to missed causal links, resulting in incorrect conclusions and wasted analyst effort.
(2) Inconsistent Adversary Simulation:
LLMs often fail to adapt as scenarios evolve, weakening simulation realism. In Resource Development (\textit{Establish Account}), adversaries pivot credentials for persistence, yet LLM-driven red teams may not escalate privileges appropriately (e.g., Abuse Elevation Control Mechanism to Session Hijacking), limiting fidelity. Without strategic adaptation, LLM simulations diverge from real-world threat behavior. While some studies use Catch the Flag (CTF) challenges to evaluate LLMs in red teaming \cite{deng2024pentestgpt, zhang2024cybench, abramovich2024enigma}, performance drops sharply beyond the easiest levels, mainly due to ``reasoning and knowledge limitations''~\cite{happe2025benchmarking}. These gaps nonetheless present new opportunities for advancing LLMs in automated cyber operations.

\subsubsection{Prompt and Tuning Sensitivity with Coverage Gaps} LLMs evidently demonstrate strong natural-language capabilities but remain highly sensitive to prompt phrasing and fine-tuning parameters, which can result in inconsistent behaviour and limited generalization.
Small changes in input or model configuration can lead to significant shifts in output, posing reliability risks. This is further compounded by limitations in information coverage during training (e.g., about specific vulnerabilities), which may collectively undermine the robustness and reliability of LLM-driven cybersecurity solutions.

\subsubsection{Evaluation Practices \& Integration Challenges} Integrating LLMs into operational cybersecurity frameworks such as Security Information and Event Management (SIEM) systems, MITRE ATT\&CK, or the NIST CSF presents several challenges. A key issue is the lack of real-world benchmarks and reliability measures. Most security AI evaluations are conducted in controlled environments that fail to capture the complexity of real-world live networks. This makes it difficult to assess how LLMs will perform in production environments and whether they can meaningfully support existing security workflows. Additionally, integration requires compatibility with existing tools, data formats, and protocols, an area still lacking standardization. As a result, organizations risk deploying LLMs without a clear understanding of their limitations or interoperability requirements.

\subsection{Risks of LLM-Enhanced Cyber Operations Technology}
Here, we detail the contextual risks emerging from the application of LLMs to cyber operations. For a more general overview of generative AI risks we refer the reader to~\cite{nistGenAIRisks}.




\subsubsection{Dual-Use and the Blurring of Threat Actor Boundaries.} 
Open-source tools lower the barriers to implementing strong cybersecurity measures but might also empower adversaries. Tools originally built for red or blue team operations are increasingly repurposed by threat actors to automate reconnaissance, craft phishing content, and generate evasive code~\cite{trendresearch}. This reduces both the skill and resource requirements to execute sophisticated attacks, especially when combined with powerful, openly accessible LLMs lacking sufficient safeguards or oversight. Red team frameworks like Cobalt Strike, Metasploit, and PowerSploit, once intended for ethical use, have long been co-opted by state-sponsored and criminal groups. Their open-source nature and resemblance to legitimate traffic complicates detection and attribution. Critically, the integration of LLMs into these toolchains may reshape the threat landscape, enabling low-skilled actors (e.g., script-kiddies) to operate with sophistication once reserved for advanced persistent threats. This convergence blurs traditional distinctions between amateur and state-funded adversaries, challenging existing threat models. CrowdStrike’s 2024 Global Threat Report~\cite{crowdstrikeReport} highlights this trend, noting the group \textit{Scattered Spider} likely used an LLM to generate PowerShell scripts in a 2023 attack. Without clear boundaries and responsible practices, ethical deployments risk accelerating the proliferation of offensive cyber capabilities.

\subsubsection{Over-reliance on Automation} 
Dependence on LLMs in security operations risks diminishing human awareness, oversight, and judgment.
As LLMs become integrated into automated pipelines, there is a growing tendency to rely on their outputs without sufficient human validation. In high-stakes scenarios this could lead to missed threats, misinterpretations, or overconfidence in flawed assessments---particularly where human security experts are sidelined from decision loops. Running weakly or even entirely non-audited code increases the risks of both unintended consequences and introducing new vulnerabilities. Analysts' ability to find and fix code vulnerabilities may be hampered by greater unfamiliarity, and there is a risk that the overall quality of code supporting business activities drifts lower over time.

More subtly, prolonged reliance on LLMs may erode the expertise and situational awareness of security analysts. This may ultimately result in a detrimental incapacity among security teams to respond effectively in the event of LLM unavailability, whether due to deliberate service restrictions (e.g., imposed by governmental entities or AI companies) or unrelated operational failures.

%
\subsubsection{Under-reliance on Automation}
Conversely, failing to adopt sufficient automation can leave security teams unable to keep pace with the speed, scale, and sophistication of modern threats. As adversaries increasingly leverage AI-powered tools to craft targeted phishing campaigns, generate convincing deepfakes, or execute large-scale probing of infrastructure, manually operated defenses may simply fall short~\cite{enisa}. Security analysts overwhelmed by alert fatigue, repetitive triage, and log analysis may miss high-impact events or respond too slowly. Underutilization of LLMs for automating routine tasks, synthesizing intelligence, or detecting anomalies not only wastes a critical force multiplier but also increases the burden on already stretched human teams. Without strategic integration of AI-driven tools, organizations risk a growing asymmetry in cyber defense, unable to match the speed and variation of LLM-enhanced attackers.


\subsubsection{Emerging Technology Risk} 

A potential risk emerging from an increased pace of technological progress, driven at large by LLM augmentation, is a widening gap between the literature and practice-of cyber security. Both attackers and defenders now benefiting from LLM-enhancement quickly move into uncharted and undocumented territory as they find new ways to apply and integrate emerging AI technologies. This leaves a significant risk of "learning the hard way" without established, or up-to-date, best-practice guidelines available. 


\subsubsection{Privacy risks}
To access the most effective LLMs users typically need to rely on API access to proprietary models. Reliance on proprietary models raises significant privacy concerns as it necessitates sharing the sensitive information for operating LLM-based agents in red and blue team scenarios with the model owners.
Even if AI companies provide privacy-preserving solutions for their customers, security vulnerabilities in the inference pipeline could still expose critical information to third parties.
Moreover, many organizations have policies that restrict sensitive data from leaving their network. Thus, even if proprietary models deliver substantial performance improvements for red and blue team operations, privacy risks may prevent customers from using them for critical cybersecurity tasks. This situation could leave organizations at a disadvantage by not fully leveraging the potential of LLMs in their cyber operations. Conversely, the growing capabilities of LLMs might tempt individual employees to use proprietary models despite potential non-compliance with their organizations data protection policies. 

\subsubsection{Agentic LLMs}
Agentic LLMs are characterized by tool use and multi-agent interactions, enabling standard LLMs to achieve more complex goals with minimal human intervention~\cite{durante2024agent,shinn2023reflexion, wu2024autogen}. In this process, many of the risks posed by standard LLMs are both exacerbated and made more challenging to estimate. The risks of over and under-reliance on LLMs both rise as agentic workflows enable tackling more complex tasks (e.g., end-to-end cyber kill chains) with less human intervention, in turn elevating the pace of threat escalation and magnifying the risks of solutions unable to move at machine speed. Tool-equipped LLMs with API access increase the worst-case risks of LLMs and heighten the privacy risks as control over which information is shared is handed off to the agent. Finally, agentic LLM systems have greater attack surfaces and pose correspondingly enlarged risks. Agentic platforms have already been found to contain exploitable vulnerabilities from which credentials could be leaked by an attacker~\cite{CVE-2025-31491}.

\section{Recommendations, and Future Directions for LLM Adoption and Readiness in Cybersecurity}


As LLMs become embedded in cybersecurity workflows, both red and blue teams face a pivotal challenge: \textit{how to harness these tools' capabilities without introducing unacceptable risks}. The integration of LLMs must be approached as a series of carefully judged trade-offs between e.g., speed and accuracy, autonomy and oversight, and innovation and safety. 
This section outlines strategic principles for responsible adoption, drawing on the technical challenges and operational risks presented earlier. 



\subsection{Mitigate Adversarial Dual-Use of LLM Technology} Highlighting the need for strong public-private coordination to detect and counter emerging threats, CrowdStrike's 2024 report~\cite{crowdstrikeReport} shows adversaries already adopting LLMs, including commercial variants, to accelerate their attacks. Open-source tools and permissive APIs further lower the technical barriers (e.g., accelerate reconnaissance and exploit development) for advanced and low-skilled actors alike. To mitigate these risks, we recommend: (1)  \textit{Develop usage constraints and API-level access control}, especially  for security-critical LLM applications. (2) \textit{Establish dual-use governance frameworks} to track and assess AI-enhanced offensive capabilities in cybersecurity, especially within open-source ecosystems. (3) \textit{Foster cross-sector collaboration} among governments, industry, and academia to establish norms and standards, and to enable timely detection and response to emerging threats, including the misuse of red team frameworks and LLM-assisted tooling.  Insights from OpenAI's research on LLM safety, Google DeepMind's robust AI systems, and initiatives by the various AI Safety and Security Institutes~\cite{bengio2025international} further indicate the value of a collaborative effort. (4) \textit{Support early warning systems} and shared intelligence platforms informed by LLM misuse trends e.g., the reported use of LLMs by threat actor group Scattered Spider in CrowdStrike's global report.

\subsection{Balance Automation with Human Oversight}  Both over- and under-reliance on automation can have adverse medium-to-long-term consequences and must be strategically balanced. While integrating LLMs to help in more tasks across incident response, threat modelling and reporting, it is critical to retain some level of human involvement and understanding 
to ensure safe deployment. 
To ensure transparency and accountability: (1) \textit{Implement human and LLMs collaboration mechanisms} especially for high-impact, ambiguous or irreversible actions. (2)  \textit{Design adaptive automation thresholds}, escalating to human intervention when model confidence is low or contextual anomalies emerge.
(3) \textit{Log decisions and generate explanations} for actions taken to support forensic analysis and compliance. 
(4) \textit{Design interactive, real-time interpretability tools} tailored to the needs of blue team operators.
These principles align with NIST’s AI Risk Management Framework and the broader effort towards safe and trustworthy AI.

\subsection{Ensure Privacy-Conscious and Secure Deployment} 

Integrating LLMs into security operations demands a privacy-conscious and security-first approach. Many deployments rely on proprietary APIs, raising risks around data exposure, compliance, and vendor lock-in. To mitigate these risks: 
(1) \textit{Sandbox and isolate deployments:} Run LLM agents in contained environments to prevent unauthorised access or unintended interactions. 
(2) \textit{ Apply least  privilege principle:} Restrict model permissions to only what is necessary for the task. 
(3) \textit{ Automate monitoring and logging:} Continuously observe model outputs and tool interactions to detect anomalies, misuse, or drift over time. 
(4) \textit{ Avoid public APIs where sensitive threat data or logs are involved} and advocate for on-premise or private inference solutions when feasible.
These measures align with the Zero Trust Architecture principles advocated by NIST and help reduce both accidental leakage and adversarial abuse.

\subsection{Strengthen Guardrails for Agentic LLMs}
Agentic LLMs with tool-use capabilities present significant benefits but also entail compound risks due to autonomy, unpredictable misbehaviour, and extended access to systems. To mitigate these risks: 
(1) \textit{Design tool access brokers} that gate and log each external API or system call made by an agentic LLM for auditibilty when needed. 
(2) \textit{Apply limiting policies} for task chaining, especially in scenarios where agents attempt multi-step operations without human review.
(3) \textit{Simulate fail-closed scenarios} where agentic systems lose access to certain tools or contexts to evaluate response robustness and investigate their unpredicted behaviour.
(4) \textit{Require red-teaming and adversarial testing} of agentic LLMs prior to deployment, focusing on unintended tool use, privilege escalation, and emergent behaviors under real-world constraints.

\subsection{Build Real-World Benchmarks to Advance Safe LLM Use and Risk Awareness} The real-world impact of LLM-driven cybersecurity research is often constrained by the lack of realistic benchmarks. Existing methods such as CTF challenges and synthetic datasets~\cite{yang2023intercode,zhang2024cybench} fail to capture the complexity and operational nuance of real-world attack and defense scenarios.
To responsibly integrate LLMs into red and blue team workflows we should also develop evaluation practices that both validate practical effectiveness, and expose limitations and risk thresholds. This can ensure these tools are not only useful but also deployed with awareness of their boundaries, enabling timely mitigation when needed. To address this gap, we recommend: 
(1) \textit{Domain-Relevant Benchmarking: }Develop evaluations grounded in real or high-fidelity environments that reflect realistic threat models and operational demands. 
(2) \textit{Community-Supported Testing Grounds:} Establish open benchmarking frameworks similar to MITRE ATT\&CK or MISP, jointly developed by cybersecurity and AI communities. 
(3) \textit{Transparent Reporting and Reproducibility:} Encourage the release of evaluation results with code, prompts, and configurations to enable replication, validation, and comparison across settings. 
(4) \textit{Practical, impact-driven metrics:} Draw on lessons from initiatives like Microsoft's AI Safety Research and the AI Incident Database (Partnership on AI), to develop metrics focused on operational performance, not just academic proof-of-concept.





\subsection{Design for Real-Time Adaptability and Agility} 
The threat landscape evolves rapidly and so must the tools used to defend against it. Effective use of LLMs in cyber operations requires agility, which involves: 
(1)~\textit{Live model updating:} Regularly feed the latest threat intelligence into LLMs to maintain relevance. 
(2)~\textit{Incorporate continuous learning pipelines} that allow for safe, rapid model updates.
(3)~\textit{Enable contextual adaptation} tailoring responses to current attack vectors and threat landscapes.
(4)~\textit{Adopt MLOps-inspired monitoring frameworks} for operational oversight and rollback when needed.

\subsection{Promote Responsible Access While Avoiding Asymmetric Advantage}
A healthy cyber ecosystem requires enabling defenders while preventing malicious actors from exploiting LLMs. To that end, we recommend. 
(1) \textit{Encourage responsible access tiers: } broad access for safe use cases (e.g., threat detection) vs. stricter review for dual-use or offensive capabilities.
(2) \textit{Incentivize LLM-for-good initiatives:} (e.g., Blue Team Olympiads, AI4Cyber Defence Grand Challenges) that build capacity in under-resourced teams.
(3) \textit{Disincentivise irresponsible model hosting and deployment:} i.e., those failing to implement misuse prevention, especially in high-risk domains.
(4) \textit{Promote open, secure access pathways for public-good applications: } for example providing vetted LLM access for nonprofit, educational, and public-sector cybersecurity efforts, ensuring these groups are not disadvantaged by limited resources while still upholding strong safeguards against misuse.

\section{Remarks}
LLMs are reshaping the cybersecurity battlefield, offering new opportunities while blurring traditional lines between defense and offense. This paper outlines the evolving role of LLMs in red and blue teaming, alongside the technical challenges and risks they introduce. As adoption accelerates, our ability to govern, benchmark, and secure these tools will determine whether they become assets for resilience or vectors for risk. We advocate for responsible and secure deployment rooted in human oversight, open collaboration, and continuous adaptation---a vision that must guide cybersecurity innovation in the age of LLMs.

 
\bibliographystyle{ACM-Reference-Format}
\bibliography{ref}
\end{document}